\documentclass[sigplan,twocolumn]{acmart}

\usepackage{quantikz}
\usepackage{amsmath}

\usepackage{algorithm}
\usepackage{algorithmic}
\usepackage[normalem]{ulem}
\usepackage{graphicx}
\usepackage{subcaption}

\newcommand{\sol}{EC2S}

\begin{document}
\title{Efficient Circuit Cutting and Scheduling in a Multi-Node Quantum System with Dynamic EPR Pairs}

\author{Zefan Du}
\affiliation{%
  \institution{Fordham University}
  \city{New York, NY}
  \country{USA}
}

\author{Wenrui Zhang}
\affiliation{%
  \institution{New Jersey Institute of Technology}
  \city{Newark, NJ}
  \country{USA}
}



\author{Wenqi Wei, Juntao Chen}
\affiliation{%
  \institution{Fordham University}
  \city{New York, NY}
  \country{USA}
}

\author{Tao Han}
\affiliation{%
  \institution{New Jersey Institute of Technology}
  \city{Newark, NJ}
  \country{USA}
}

\author{Zhiding Liang}
\affiliation{%
  \institution{Rensselaer Polytechnic Institute}
  \city{Troy, NY }
  \country{USA}
}

\author{Ying Mao}
\affiliation{%
  \institution{Fordham University}
  \city{New York, NY}
  \country{USA}
}\thanks{Corresponding authors: \{zdu19, ymao41\}@fordham.edu}

\begin{abstract}

Despite advancements, current quantum hardware faces significant challenges, including limited qubit counts and high susceptibility to noise, which hinder the execution of large, complex algorithms. To address these limitations, multi-node quantum systems and quantum circuit cutting techniques partition large circuits into smaller subcircuits that can be executed on individual quantum machines and then reconstructed using classical resources. However, these methods introduce new challenges, such as the large overhead from subcircuit reconstruction and additional noise from entangled EPR pairs, especially in multi-node quantum systems. In this paper, we propose the Efficient Circuit Cutting and Scheduling (\sol) system, which integrates EPR pairs with circuit cutting to address these issues. \sol~ improves system performance by transitioning from logical to physical EPR pairs and further reduces computational overhead by minimizing the number of subcircuits during the reconstruction phase. \sol~ is implemented using Qiskit and evaluated on both real quantum hardware and various emulators. Compared to the state-of-the-art Qiskit-Addon-Cut, \sol~ achieves significant improvements in fidelity, up to 16.7\%, and reduces system-wide expenditure by up to 99.5\%.


    
\end{abstract}


\maketitle 

\section{Introduction}

Quantum computing has garnered significant attention due to its potential to solve problems intractable for classical computers. Foundational algorithms like Shor’s algorithm~\cite{shor1994algorithms} for integer factorization and Grover’s algorithm~\cite{grover1996fast} for unstructured search have demonstrated its theoretical capabilities. Recently, tech giants such as IBM, Google, Microsoft, Amazon, and NVIDIA have begun offering quantum or quantum-classical cloud services. Building on these advancements, quantum computing has been studied and increasingly applied to fields like chemical simulation~\cite{mcardle2023quantum,tacchino2023quantum,honeychurch2022quantum}, machine learning~\cite{carleo2022machine,hoang2022efficient, mu2022iterative,liu2022quantum,zhang2023quantum,labbate2024quantum, 9605352, stein2022quclassi, stein2021hybrid, stein2022qucnn, jiang2024resource}, quantum encryption~\cite{honjo2022long,cao2023high,alagic2022status}, and visualization~\cite{ruan2023quantumeyes, spivak2023vmd, ruan2023venus, ruan2022vacsen}.

Despite its immense potential, current quantum hardware is still in its early stages and faces significant challenges, notably limited qubit counts~\cite{fujii2023quantum} and high susceptibility to noise. Various physical qubit technologies—superconducting qubits, trapped ions, neutral atoms, and photonic circuits—are progressing toward commercial quantum computers but continue to suffer from a shortage of stable qubits. Moreover, noise and errors severely impact quantum fidelity, which reflects the accuracy of quantum state manipulation and measurement. Low fidelity results from quantum gate errors~\cite{epstein2022noise}, imperfect state preparations and controls~\cite{zhang2023characterizing, baheri2022pinpointing}, decoherence~\cite{rossi2022mitigating}, and qubit cross-talk~\cite{alexander2022mitigation}. Quantum error correction codes are designed to mitigate errors due to decoherence, quantum noise, and other disturbances affecting qubit quality during computation. Quantum compilers aim to map circuits to physical qubits to reduce noise effects and increase fidelity. However, as quantum systems scale, maintaining high fidelity becomes increasingly challenging. Consequently, today’s quantum devices are mostly in the NISQ (Noisy Intermediate-Scale Quantum) era with 50–100 qubits. While these systems have shown early promise with small circuits, they are not yet capable of handling the large and complex algorithms needed for practical applications.

To address the challenges of noise and scaling in executing large quantum circuits, multi-node quantum systems and quantum circuit cutting techniques have received significant attention. A multi-node quantum system connects multiple smaller quantum processors to construct a distributed system, offering a promising solution for building large-scale quantum systems without the difficulties of manufacturing a single large monolithic device. In such systems, multiple quantum processors—referred to as workers—are interconnected to surpass the computational power of a single device. EPR pairs (named after the Einstein-Podolsky-Rosen paradox) are commonly used to connect these workers, allowing non-local gate operations like CNOT (CX) gates to be executed remotely~\cite{briegel1998quantum, dur1999quantum}. By employing quantum circuit cutting techniques, we can partition a large circuit into smaller subcircuits executable on smaller quantum machines. These subcircuits are then reconstructed using classical resources, such as CPUs or GPUs, to recover the quantum states of the original circuit. Leveraging multiple connected quantum machines of a more manageable size, this quantum-classical architecture holds the potential to build large-scale quantum systems.

However, integrating circuit cutting into a multi-node system introduces new challenges. Generating and distributing EPR pairs is subject to noise and delays, requiring careful qubit management~\cite{wang2020integrated, luo2020trapped}. The system must optimize communication paths and the placement of remote gate operations to reduce overhead and improve performance. Additionally, it needs to analyze specific circuits to identify optimal cutting points that maintain high algorithmic fidelity while reducing reconstruction costs. Scheduling subcircuits to appropriate quantum workers is necessary to mitigate noise impacts. Increasing the number of cutting points leads to more subcircuits, resulting in significant sampling overhead during the reconstruction phase.




This paper aims to address these challenges by proposing an Efficient Circuit Cutting and Scheduling (\sol) framework in a multi-node quantum system with dynamic EPR pairs. We propose \sol, a framework that manages a multi-node quantum system and accepts large quantum circuits as input. It analyzes the input circuit to find cutting points that minimize reconstruction overhead and schedules subcircuits to quantum workers to reduce noise impacts. Specifically, \sol~ dynamically generates EPR pairs to link adjacent subcircuits, virtually merging them into larger subcircuits via remote gate operations. This approach reduces the number of cut points and, consequently, lowers the reconstruction overhead. Given the success rate of generating EPR pairs, \sol~ strategically schedules subcircuits to quantum workers in a manner that minimizes noise impacts.


Our contributions include the following:
\begin{itemize}
    \item We introduce circuit cutting into a multi-node quantum system connected by EPR pairs, aimed at significantly reducing reconstruction overhead.
    
    \item We propose \sol~, a framework with a suite of noise-aware algorithms. \sol~ identifies the best subcircuits to be connected with logical EPR pairs and selects the optimal workers to dynamically generate physical EPR pairs, both in a noise-aware manner. The subcircuits are then distributed across workers to minimize overall system expenditures.
    
    \item We implement \sol~ using IBM Qiskit and Quantinuum Pytket, evaluating it on both real quantum hardware and quantum emulators. Through extensive experimentation, \sol~ outperforms the state-of-the-art Qiskit Addon Cut in a multi-node environment, achieving up to a 16.7\% improvement in fidelity and reducing overhead by as much as 99.5\%.
\end{itemize}

\section{Related Works}

Recent advances in multi-node quantum systems and quantum circuit cutting have attracted attention for addressing the challenge of executing large-scale quantum circuits on scalable hardware. ARQUIN~\cite{ang2024arquin} represents a collaborative effort among government, industry, and academia to design multi-node quantum systems based on superconducting architectures, evaluating performance in terms of internode links, entanglement distillation, and local architecture. Several protocols and applications have been proposed for these systems~\cite{caleffi2024distributed, cuomo2023optimized, d2023distributed, singh2021quantum, cao2022single, gustafson2021large}.

To execute large quantum circuits on multiple smaller quantum computers, we need to partition large circuits into smaller subcircuits. While theories for both wire cuts and gate cuts have been proposed~\cite{peng2020simulating}, CutQC~\cite{tang2021cutqc} was the first implemented solution, dividing quantum circuits into independently executable subcircuits with classical post-processing used to reconstruct the original circuit's output. However, CutQC suffers from large sampling overhead and scalability issues. Follow-up works aim to strategically cut circuits to minimize computational overhead and noise impacts~\cite{chen2024quantum, kan2024scalable, brandhofer2023optimal, anand2022quantum, lierta2022optimal, smith2023clifford}. For instance, \cite{anand2022quantum} proposes an error correction circuit cutting method that optimizes cutting points in noisy environments, while \cite{lierta2022optimal} develops optimization algorithms based on local information to quickly find optimal cutting points in quantum circuits. Recent studies~\cite{bergholm2022pennylane, alexander2022mitigation} explore gate cutting to reduce circuit complexity and mitigate noise, including combining gate cuts with quantum error correction techniques to improve computation fidelity. These works highlight that gate-level partitioning can complement or even surpass traditional wire cutting methods, especially in deep quantum circuits. Despite IBM Qiskit's development of the Qiskit Addon Cutting library~\cite{qiskit-cutting}, practical implementation remains challenging due to significant overhead and resource demands in cut point identification and reconstruction phases.

Given a multi-node system, researchers and industry are exploring efficient ways to establish quantum links between quantum processors, with EPR pairs playing a key role in enabling entanglement for distributed quantum systems. The research projects like AutoComm~\cite{9923799}, CollComm~\cite{wu2022collcommenablingefficientcollective}, and QuComm~\cite{wu2023qucomm} optimize EPR usage by reducing communication overhead, incorporating buffers, and improving inter-node gate routing, significantly enhancing fidelity and efficiency in large-scale systems. In parallel, companies like IBM~\cite{monroe2020ibm}, Google~\cite{arute2019quantum}, and Amazon~\cite{aws2021quantum} are using EPR pairs to scale distributed quantum computing, with Google leveraging them in their quantum supremacy milestone. Additionally, quantum key distribution (QKD) systems from ID~Quantique and Toshiba utilize EPR pairs for secure communication~\cite{scarani2009qkd}. Together, these advances in academia and industry are driving the scalability and practicality of distributed quantum systems.

Unlike existing approaches, we address the challenges of circuit-cutting in a multi-node quantum system by aiming to reduce reconstruction costs through the use of EPR pairs. When an input circuit is partitioned into subcircuits, the system must select which pairs of subcircuits to connect with logical EPR pairs. Additionally, the quantum workers in this multi-node system must determine where to generate physical EPR pairs. Furthermore, subcircuits need to be distributed across workers in a way that balances the noise introduced by EPR pairs with the benefits gained from them.


\section{Background}
In this section, we introduce the necessary background knowledge of \sol~ system, including the circuit cutting, EPR pairs entanglement, and noise.

\subsection{Quantum Circuit Cutting and Reconstruction}

Quantum circuit cutting is the process of partitioning a large circuit into smaller subcircuits that can be executed independently on current hardware. The results from these subcircuits are then reconstructed into the original outcome through classical post-processing~\cite{bravyi2016trading, peng2020simulating}. This technique extends the capabilities of quantum hardware with limited qubit counts. There are two types of circuit cuts: wire cuts and gate cuts. A wire cut involves severing a qubit’s path between gates, allowing the circuit to be split into segments that can be processed separately~\cite{peng2020simulating}. A gate cut, on the other hand, splits a multi-qubit gate—like a controlled-NOT (CNOT) gate—into operations that can be distributed across subcircuits. Both methods reduce circuit complexity, enabling large-scale quantum algorithms to run on limited hardware.

\begin{figure}[htbp]
    \centering
    \includegraphics[width=0.99\linewidth]{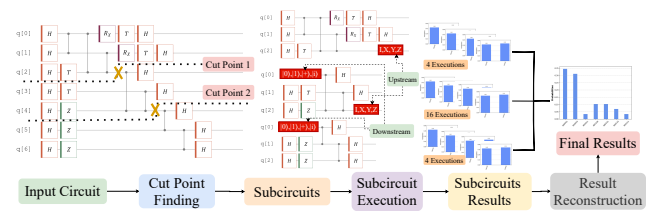}
    \caption{A circuit cutting and reconstruction pipeline.}
    \label{fig:circuit_cutting}
\end{figure}

Wire cutting is particularly valuable in hybrid quantum-classical workflows, where quantum hardware performs part of the computation and classical algorithms handle post-processing. In this project, we focus on wire cuts. 
Figure~\ref{fig:circuit_cutting} illustrates a circuit cutting example with a 7-qubit input circuit. It has been partitioned into three subcircuits with two cutting points. 


After execution, the measurement outcomes from the subcircuits are collected for classical post-processing and result reconstruction. For each cut, there are two subcircuits: upstream and downstream. The upstream subcircuits are executed with four basis states ($I, X, Y, Z$) for qubits at the cut locations, while the downstream subcircuits use four different initialization ($\ket{0}, \ket{1}, \ket{+}, \ket{i}$). The probability distribution for subcircuit 1 with basis state $i$ is denoted as $p_{1,i}$. The probability distribution for reconstructing subcircuits 1 and 2 is:

\begin{equation}
    \text{P} = \frac{1}{2} \sum_{i=1}^{4} p_{1,i} \otimes p_{2,i}
    \label{eq:1}
\end{equation}

Based on Equation~\ref{eq:1}, as illustrated on Figure~\ref{fig:circuit_cutting} Subcircuit-1 (Top) and Subcircuit-3 (Bottom) have $4$ different probability distributions. Therefore, they need to be executed 4 times. Subcircuit-2 (Middle) contains both upstream and downstream that leads the $16$ different probability distributions and requires 16 executions.
The results from the subcircuits are combined using Kronecker products to reconstruct the full probability distribution of the original circuit. This reconstruction sums the Kronecker products of the subcircuits' measurement outcomes to calculate the final results.

Reconstructing the output to the original quantum circuit requires classical post-processing to merge subcircuit results. The main challenge lies in the exponential growth of classical resources needed for post-processing as the number of cuts increases. Each subcircuit's measurement results are taken across different basis states and combined through Kronecker products. As the number of cuts rises, the classical overhead can become a computational bottleneck.




\subsection{Einstein-Podolsky-Rosen (EPR) pair}
Remote Einstein-Podolsky-Rosen (EPR) pair is one of the most well-know entangled states as inter-node quantum communication. It is fundamental for enabling non-local quantum gates~\cite{einstein1935can}. A remote EPR pair represents an entangled state between two qubits residing on different compute nodes, expressed as $\frac{\ket{00}+\ket{11}}{\sqrt{2}}$. This pair acts as a quantum communication channel, facilitating quantum data transfer between nodes.

The multi-node quantum system architecture relies on remote EPR entanglement, allowing nodes in the quantum network to form various topologies through the configuration of these entangled connections. On a specific node, not all physical qubits are capable of establishing remote EPR entanglement ~\cite{andres2019automated}. The compute node also contains data qubits, which are used to store program information. This architecture allows for flexible quantum network designs through the strategic use of remote EPR entanglement.

\begin{figure}
    \centering
    \includegraphics[width=1\linewidth]{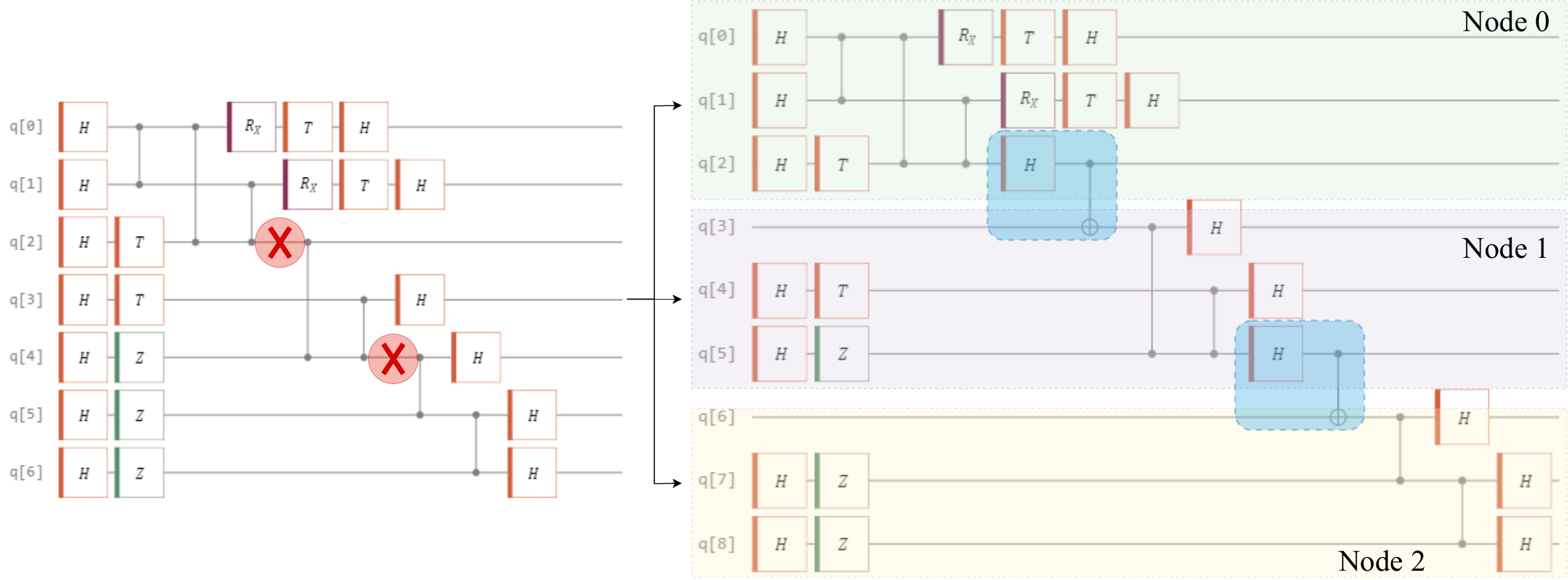}
    \caption{A Circuit Cutting Example in a 3-node quantum system connected with two EPR Pairs}
    \label{fig:2epr}
\end{figure}

Figure~\ref{fig:2epr} depicts the process of generating two pairs of EPR states between three quantum computer nodes. The EPR pair is created by applying a Hadamard (H) gate to the first qubit, placing it in a superposition, followed by a controlled-NOT (CNOT) gate, which entangles the two qubits. In the subcircuit of node 1 (blue), qubits 0-2 from the original circuit are present, along with an additional qubit (qubit 3) used for the EPR pair shared between node 1 and node 2. The subcircuit of node 2 (purple) contains the information of qubits 3-4 from the original circuit, plus an extra qubit (qubit 6) for the EPR pair shared between node 2 and node 3. Node 3 (yellow) holds the subcircuit for qubits 5-6 from the original circuit.

\subsubsection{Hardware Implementation}
Recent hardware implementations are advancing the capabilities of distributed quantum computing. For instance, recent developments in superconducting qubit systems have demonstrated high-fidelity EPR pair generation, achieving success rates above 90\% in controlled experiments \cite{song2019generation}. Photonic systems are also making progress, with advances in integrated quantum photonics improving entanglement distribution rates and scalability \cite{wang2020integrated}. Moreover, trapped-ion systems have shown reliable EPR generation with success rates improving through error correction protocols \cite{luo2020trapped}. Despite these advances, noise remains a key limiting factor in long-distance distributed quantum computing, making further improvements in error correction, purification, and quantum repeaters crucial for enhancing the scalability and reliability of EPR-based quantum networks \cite{luo2020trapped,wang2020integrated}.

\subsection{Noise}
Noise is one of the most significant challenges in quantum computing, as quantum states are highly sensitive to interactions with their environment. Even small perturbations can induce errors that degrade the performance of quantum algorithms. This section discusses noise relevant to quantum systems, with a focus on gate noise, noise affecting EPR pairs, and common noise models used to simulate quantum circuits.

\subsubsection{Depolarizing Noise Model}
In the depolarizing noise model, a qubit undergoes a random Pauli error with probability \( p \). The depolarizing channel acting on a single qubit state \( \rho \) is defined as:
\begin{equation}
    \mathcal{E}_{\text{depolarizing}}(\rho) = (1 - p) \rho + \frac{p}{3} \left( X \rho X + Y \rho Y + Z \rho Z \right),
    \label{eq:depolarizing_noise_single}
\end{equation}
where \( X \), \( Y \), and \( Z \) are the Pauli matrices~\cite{nielsen_chuang}. 

For single-qubit gates, depolarizing noise causes the qubit to deviate from its intended trajectory by applying one of the Pauli operators randomly, thereby reducing the fidelity of the computation. 
For two-qubit gates, the depolarizing noise model is extended to account for errors on both qubits simultaneously. The action of the two-qubit depolarizing channel on a two-qubit density matrix \( \rho_{AB} \) can be written as:
\begin{equation}
    \mathcal{E}_{\text{depolarizing}}(\rho_{AB}) = (1 - p) \rho_{AB} + \frac{p}{15} \sum_{i,j} P_i^A P_j^B \rho_{AB} P_i^A P_j^B,
    \label{eq:depolarizing_noise_twoqubit}
\end{equation}

where \( P_i^A \) and \( P_j^B \) are Pauli operators acting on qubit \( A \) and qubit \( B \), respectively, and the summation is over all possible combinations of Pauli matrices on both qubits. 

Two-qubit gates are generally more prone to depolarizing noise than single-qubit gates because of the larger number of possible error configurations, and the error rate \( p \) is typically higher for two-qubit gates due to the increased complexity of implementing interactions between qubits. 
This increased error rate necessitates careful error mitigation and quantum error correction strategies, especially in circuits with many two-qubit operations, as the accumulation of depolarizing noise can significantly degrade the overall fidelity of the quantum computation~\cite{bravyi2016trading}.

\subsubsection{Noise for EPR Pairs}
EPR pairs, or Bell states, are maximally entangled states of two qubits and are commonly used in quantum communication to enable remote gate operations in multi-node quantum systems. These pairs are generated dynamically between two quantum nodes but are highly susceptible to noise. Major sources of noise include decoherence, photon loss, and gate errors, all of which reduce the fidelity of EPR pairs. Photon loss is particularly problematic in optical quantum communication, where the success rate of EPR pair distribution decreases exponentially with distance, modeled by \(P_{\text{success}} = e^{-\alpha d}\), where \( \alpha \) is the loss coefficient and $d$ is the distance. In typical optical fibers, the loss coefficient is around 0.2 dB/km, which drastically reduces the success rate for distances exceeding 100 km \cite{briegel1998quantum}. In photonic systems, the success rate for generating and detecting entangled photons typically ranges between 1\% and 10\%~\cite{slussarenko2017photonic}. 



In short-distance or localized quantum systems (e.g., superconducting qubits, trapped ions and neutral atoms), the success rate is typically modeled as a constant, assuming high-fidelity gates with near-deterministic entanglement generation. These hardware architectures offer much higher success rates, often exceeding 90\% for short-distance entanglement~\cite{song2019generation}. However, noise remains a key limiting factor in long-distance distributed quantum computing. Advances in error correction, entanglement purification, and quantum repeaters are essential to improving the scalability and reliability of EPR-based quantum networks.

\section{Problem Formulation}

In this project, we aim to utilize EPR pairs in a multi-node quantum-classical system to implement circuit cutting techniques. It facilitates the execution of larger quantum circuits while minimizing overall system costs. 


\begin{figure}
    \centering
    \includegraphics[width=1\linewidth]{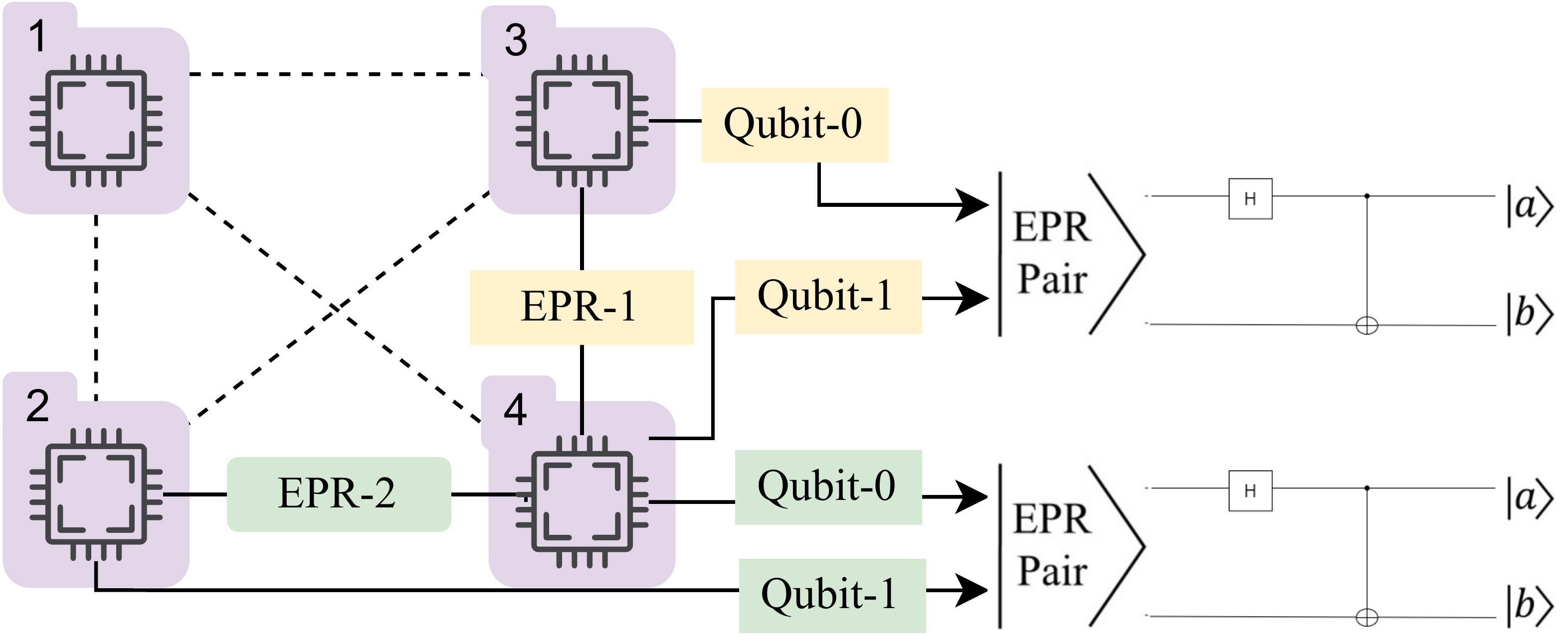}
    \caption{A 4-node system with 2 EPR Pairs}
    \label{fig:topo}
\end{figure}

\subsection{Problem Setting}
We consider an all-to-all EPR-connected multi-node quantum system, where EPR pairs are dynamically generated but subject to noise, which reduces fidelity. Figure~\ref{fig:topo} illustrates a 4-node system with potential connections between any two nodes (dashed lines). In this example, EPR pairs are generated between Nodes 2 and 4, and Nodes 3 and 4, with a physical resource cost of 1 qubit on Nodes 2 and 3, and 2 qubits on Node 4.




Formally, the system consists of \(n\) quantum workers \(W = \{w_1, \dots, w_n\}\), each with a specific qubit count \(Q = \{Q_1, \dots, Q_n\}\) and an associated noise model \(N = \{N_1, \dots, N_n\}\). Workers can be dynamically connected using EPR pairs, denoted by \(E\), with up to \(\binom{n}{2}\) possible pairs in an \(n\)-node system, and each pair has a success rate \(SR\). While increasing the number of EPR pairs can reduce reconstruction costs, it also introduces more noise, lowering the overall algorithmic fidelity.





We investigate the problem of circuit cutting and reconstruction in such a system. The input circuit \(C\) can be partitioned into \(m\) subcircuits, represented as the set \(S = \{s_1, \dots, s_m\}\). Given the multi-node system with \(W\) workers and up to \(E\) EPR pairs, we focus on three main objectives: (1) identifying pairs of subcircuits, \(s_i\) and \(s_j\), that can be connected with EPR pairs based on the structure and cutting points of the subcircuits, and their serial numbers will be merged, and the serial numbers of the subcircuits after the back sorting number will be reduced by one (i.e., \(i < j < m\) , \(s_i + s_j \Longrightarrow s_{ij}\) , \(s_m \Longrightarrow s_{m-1}\) ).; (2) reconfiguring the set of workers \( W = \{w_1, \dots, w_n\} \) to find the optimal pair \(w_i\) and \(w_j\) for generating EPR pairs, considering the noise models \(N\); and (3) scheduling the subcircuits \(s_1, \dots, s_{m-e}\) to the quantum workers \(w_1, \dots, w_n\), taking into account EPR-connected pairs \((w_i, w_j)\) and their associated noise.

\subsection{Objectives and Constraints}
In addition to these objectives, we aim to optimize the system using a metric, Multi-node System Expenditure (MSE), which considers sampling overhead, noise from EPR pairs, and total resource requirements.


\noindent\textbf{Sampling overhead} refers to the additional computational and experimental cost incurred to calculate the probability distribution of quantum states after multiple executions (or "shots") of the same circuit. The required number of shots increases with the number of qubits and error rates, impacting the overall computational load.
In a multi-node quantum system, where circuits are partitioned into subcircuits, the total sampling overhead is the sum of the overheads for all subcircuits. When subcircuits are interconnected with EPR pairs, the required number of shots increases by a factor dependent on the EPR success rate \(SR\), represented as \( (1 - SR) \), where \(SR \subset [0, 1]\). This accounts for the additional uncertainty from potentially unsuccessful EPR operations, requiring more measurements (i.e., shots) to maintain accuracy. Additionally, based on the noise model $N$, we use $\epsilon$ to determine the error rate of the subcircuit $s_i$.

The sampling overhead \( O(S) \) for a quantum circuit, $C$, that is partitioned into $m$ subcircuits $\{s_1,...,s_m\} \subseteq S$,  can be defined as: 

\begin{equation}
    O(S) = \prod_{i=1}^{m} \frac{1}{(1 - 2\epsilon_j(s_i))^2} \times N_{\text{shots}} (s_i) 
    \label{eq:overhead}
\end{equation}
Where:
\begin{itemize}
  \item \( \epsilon_j(s_i) \) is the error rate for the \( i \)-th subcircuit \( s_i \), potentially varying by subcircuit due to different worker \(W_j\) with different types of operations, and worker capabilities.
  \item \( N_{\text{shots}} (s_i) \) is the number of shots required to execute the \( i \)-th subcircuit \( s_i \) with adequate statistical precision.
  \item \( m \) is the number of subcircuits resulting from the circuit cutting.
\end{itemize}


Using Equation~\ref{eq:overhead}, we assume that higher error rates significantly increase the number of shots needed to accurately estimate the quantum system's state. Our goal is to find optimal cuts that minimize the sampling overhead for each subcircuit and strategically select subcircuit pairs \((s_i, s_j)\) to be connected via EPR pairs. Once the subcircuits are connected via EPR pairs, we use the subcircuit $s_{ij}$ with the larger number of qubits between the two to calculate the overhead.
When introducing EPR pairs into the system, the total overhead after assigning EPR pairs is written as:

\begin{equation}
\begin{split}
    O_{EPR}(S) = \prod_{i=1}^{m-e} \frac{1}{(1 - 2\epsilon_j(s_i))^2} \times N_{\text{shots}} (s_i) + \\ \prod_{i=1}^{e} \frac{1}{(1 - 2\epsilon_j(s_{ij}))^2} \times N_{\text{shots}} (s_{ij})  \times (1 - SR)
\end{split}
\label{eq:epr_overhead}
\end{equation}

Where:
\begin{itemize}
  \item \( e \) is the number of EPR pairs used to connect subcircuits.
  \item \( SR \) is the success rate of generating and utilizing EPR pairs to connect the subcircuits, impacting the effective overall operation fidelity.
  \item \(s_{ij}\) is the subcircuit with larger qubits of two subciruit $s_i$ and $s_j$.
\end{itemize}


Equation~\ref{eq:epr_overhead} calculates the sampling overhead, considering the complexities introduced by circuit partitioning, error rates, and the impacts of using EPR pairs to connect subcircuits.


\noindent\textbf{Total Resources Requirement:} 
One of the main challenges of circuit cutting techniques is the significant sampling overhead and high resource demands they require. These resources can be divided into two categories: (1) Quantum resources, which are required for executing each subcircuit and include qubits and entanglement resources like EPR pairs. In this case, sampling overhead refers to the quantum device resources used for executing subcircuits. (2) Classical resources, which encompass the computational power needed to process and reconstruct quantum information from multiple subcircuits. This involves handling large matrices representing circuit states, and as shown in Equation~\ref{eq:1}, the demand for classical resources grows exponentially, posing further scalability challenges for circuit cutting techniques. In our system, we model the  System Expenditure in Muti-node environment (SEME) as follows:

\begin{equation}
    SEME = 4^{K- 2e} \prod_{ i = 1}^{m - e} 2^{sq_{i}} \times O_{EPR}(s_i)
    \label{eq:mse}
\end{equation}
\label{eq:SEME}
\begin{itemize}
    \item $K$: The number of cuts. For each optimal wire cutting, $K$ is set to 2.
    \item $sq_i$: The number of qubits in subcircuit $s_i$.
    \item $O_{EPR}(s_i)$: The sampling overhead of subcircuit $s_i$
\end{itemize}


As the number of EPR pairs used ($e$) increases, the reconstruction overhead decreases, reducing the overall computational cost of reconstructing subcircuits. However, they also introduce noise, decreasing fidelity. Therefore, minimizing resource usage requires carefully selecting the number of EPR pairs $e$, taking into account the success rate ($SR$), overhead reduction and error accumulation. Finding the optimal balance between the number of EPR pairs and the system's fidelity is essential.

\section{Solution Design}

In this section, we present the design of the \sol~ system and describe its key algorithms in detail.

\subsection{System overview}

\begin{figure*}[htbp]
\centering
\includegraphics[width=0.95\linewidth]{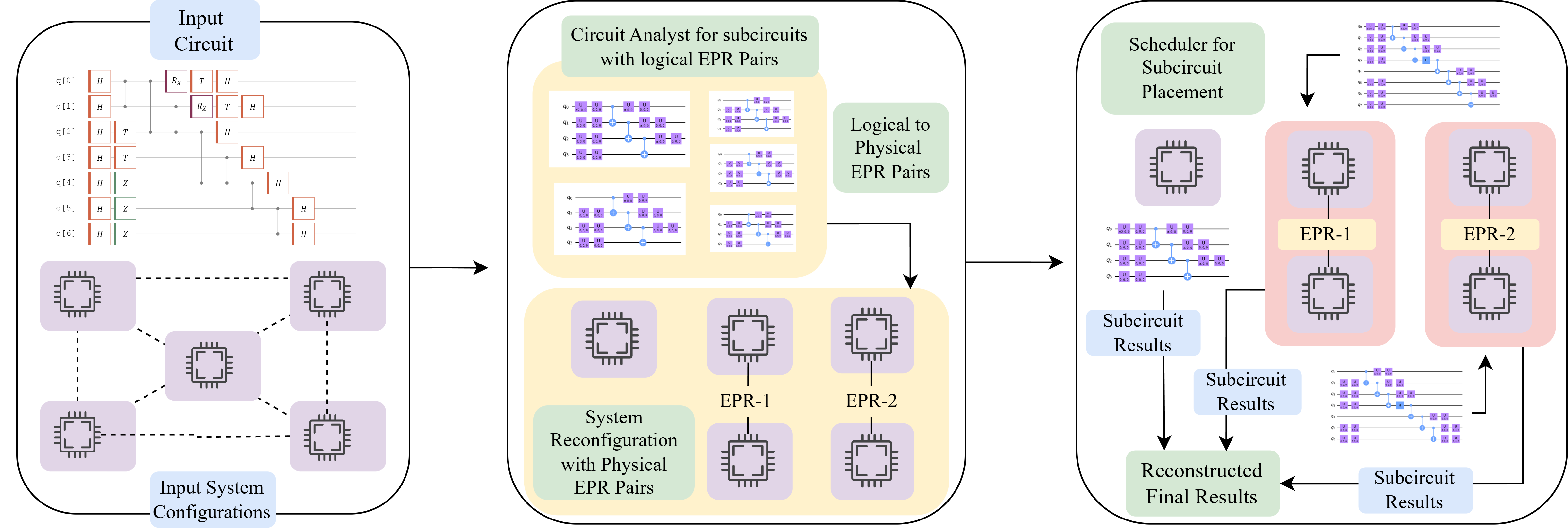}
\caption{EC2S System Overview}
\label{fig:overview}
\end{figure*}


Figure~\ref{fig:overview} provides an overview of the system. \sol~ is a multi-node quantum system that accepts two inputs: large quantum circuits from the client side (e.g., from Qiskit or Pennylane) and system hardware configurations from the administrator (e.g., number of workers, qubit counts, and noise information). Typically, the input circuit is too large to fit on a single quantum worker, and the workers are dynamically connected via EPR pairs on an as-needed basis. While using more EPR pairs reduces overhead, it also introduces more noise, lowering fidelity. \sol~ aims to find the balance between them.

The input circuit is first processed by the \textbf{Circuit Analyst} module, which partitions the circuit into subcircuits and identifies the best subcircuit pairs to be connected via an EPR pair. For example, as shown in the figure~\ref{fig:2epr}, the input circuit is divided into five subcircuits, \(s_1, \dots, s_5\). \sol~ aims to find one or more subcircuit pairs, such as \((s_i, s_j)\), and combine them to form larger subcircuits, with the goal of minimizing sampling overhead, as outlined in Equation~\ref{eq:overhead} (detailed in Algorithm~\ref{alg:1}). The analyst then virtually connects the subcircuits with logical EPR pairs.


Once logical EPR connections are established, the next step is to identify the best quantum worker pairs, \(w_i\) and \(w_j\), to physically connect by generating EPR pairs, with one qubit from \(w_i\) and another from \(w_j\). This is handled by the \textbf{System Reconfiguration} module. During this process, \sol~ evaluates different combinations of worker pairs to minimize total sampling overhead, as described in Equation~\ref{eq:epr_overhead}. The system iterates through these combinations to find the optimal number of EPR pairs and the corresponding worker pair assignments (detailed in Algorithm~\ref{alg:2}).



With the subcircuits and quantum workers configured, the \textbf{Scheduler} module assigns subcircuits and logically connected subcircuit pairs to specific workers. The scheduler takes into account each worker’s noise model, using Equation~\ref{eq:depolarizing_noise_single} for single-qubit gates and Equation~\ref{eq:depolarizing_noise_twoqubit} for two-qubit gates. To minimize error rates, subcircuits with the fewest gates affected by noise are prioritized for optimal worker assignment, thus reducing the overall noise impact (detailed in Algorithm~\ref{alg:3}).



Once scheduling is complete, the subcircuits are executed on their assigned workers. The outcomes are then collected and used in the reconstruction phase to compute the original circuit's output through post-processing on classical computing resources. \sol~ is designed to maintain the fidelity of quantum circuit execution while minimizing the consumption of both quantum and classical resources.

\subsection{Algorithms Design}
In this subsection, we present a robust approach for optimizing the partitioning and scheduling of quantum subcircuits, specifically tailored for EPR-based circuit cutting in distributed quantum systems. First, \sol~ utilizes Algorithm~\ref{alg:1} to process subcircuits and select the optimal pair that minimizes the overhead, denoted as $O(S)$. Additionally, given the set of subcircuits, Algorithm~\ref{alg:2} is designed to find the best worker pairs to minimize the EPR-based overhead, $O_{EPR}(S)$. Next, Algorithm~\ref{alg:3} schedules the subcircuits with EPR pairs onto selected workers, aiming to minimize the total resource requirement, measured as the $SEME$ cost, as defined in Equation~\ref{eq:mse}. Finally, Algorithm~\ref{alg:4} integrates Algorithms~\ref{alg:1}-\ref{alg:3} to partition the circuits, execute the cut subcircuits, and efficiently reconstruct the original circuit from the results.

\begin{algorithm}
\caption{Circuit Analyst: Logical EPR on Subcircuits}
\begin{algorithmic}[1]
\STATE Input: 
\item[] $S=\{s_1:sq_1, s_2:sq_2,...,s_i:sq_i\}$ : All subcircuits, subcircuit $s_i$ have $sq_i$ qubits
\item[] $e$ : Number of EPR pairs
\STATE Output:
\item[] $SE = [(s_i,s_j),...]$ : subcircuit pair 
\STATE MS = [ ]
\FOR{$S$}
    \STATE $SQ \gets s_i \gets min[O(s_i), O(s_{i+1})]$
    \STATE $MS.append(SQ)$
\ENDFOR
\STATE $MS.sort(key=\lambda x: x[1])$
\STATE $MI$ = [item[0] for item in MS]
\FOR{$e$}
    \STATE $SE.append(MI.pop())$
\ENDFOR
\RETURN $SE$
\end{algorithmic}
\label{alg:1}
\end{algorithm}

\subsubsection{Circuit Analyst: Identify Subcircuit Pairs}
Algorithm~\ref{alg:1} designed to find multiple pairs of subcircuits that utilize EPR pairs for quantum communication between subcircuits. It processes all available subcircuits based on the number of qubits they contain, adjusting for EPR pair connections. The algorithm~\ref{alg:1}accepts two inputs: All subcircuits dictionary $S=\{s_1:sq_1, s_2:sq_2,...,s_i:sq_i\}$ with index $s_i$ and number of qubits $sq_i$ that $s_i$ contains, and  Number of Epr pairs $e$. The output is a list, $SE$, which contains the pairs of subcircuits that are assigned EPR pairs.

The algorithm initializes an empty list, $MS$. This list is used to store the subcircuits along with the updated qubit count, including one additional qubit for each EPR pair assignment. This step is crucial as the EPR pairs consume resources(Line 3). Next, The algorithm iterates over all subcircuits $S$. For each subcircuit $s_i$, we aim to find the minimize overhead from equation~\ref{eq:overhead}, it creates a new tuple, \textbf{SQ}, which contains the subcircuit index $s_i$ and the total qubits after including the qubit needed for EPR pairing. This tuple is then appended to $MS$(Line 4-7). 
After the loop, the $MS$ list is sorted in ascending order based on the total number of qubits for each EPR pair(Line 8). 
Once $MS$ is sorted, the subcircuit indices are extracted into a new list, \textbf{MI}, which contains only the subcircuit pairs indices, the sort is from the smallest sum of qubit to the largest, that represent the reducing the minimum value of sampling overhead to the maximum value.(Line 9). 
The algorithm iterates $e$ times, corresponding to the number of available EPR pairs. In each iteration, it pop the index of subcircuit with the most qubits for EPR pair at the end of the $MI$ list. The index of this subcircuit is appended to the $SE$ list, which stores the pair of subcircuits assigned EPR pairs(Line 10-12).
Finally, the list $SE$ is returned as the output, containing the indices of the subcircuits that have been assigned EPR pairs(Line 13).

\begin{algorithm}
\caption{System Reconfiguration: Physical EPR on Workers}
\begin{algorithmic}[1]
\STATE Input: 
\item[] $C$ : circuit
\item[] $Q, E \in W$: system configurations 
\STATE $SC \& e \gets FM(C, W)$
\STATE $SE \gets Algorithm~\ref{alg:1}(S, E)$
\FOR{$SE$}
    \STATE $S_{EPR} \gets AE(S\_SE)$
\ENDFOR
\RETURN $S_{EPR}$
\end{algorithmic}
\label{alg:2}
\end{algorithm}

\subsubsection{System Reconfiguration: Find Worker Pairs}
After executing Algorithm~\ref{alg:1},  the generated subcircuit pair will be integrated into the system by identifying a pair of workers for establishing logical EPR connections. Algorithm~\ref{alg:2} is designed to find subcircuits within a larger quantum qubits that can be connected using EPR pairs in worker pair $(w_i,w_j)$. The algorithm takes three inputs: $C$: The original quantum circuit that needs to be partitioned. $E$: The number of EPR pairs in system available to establish communication between subcircuits. $Q$: The number of qubits available in the quantum processing worker. The algorithm invokes \textbf{FM} a function on finding from the circuit $C$, the number of qubits $Q$ on node, and the EPR $E$, a set of subcircuits, $SC$ and $e$ chosen for algorithm~\ref{alg:1}(Line 2). Then, the algorithm calls the algorithm~\ref{alg:1} by passing these two results. The output of this function is $SE$, a list of subcircuits pairs for connecting with EPR pairs(Line 3). The algorithm iterates through each subcircuit pair in the $SE$ list. For each subcircuit pair $S\_SE$, adds EPR connections between the subcircuits by \textbf{AE} adding EPR function, that shows a example in figure~\ref{fig:2epr} (Line 4-6). Eventually, the algorithm returns the updated cutted subcircuits $S_{EPR}$(Line 7).

\begin{algorithm}
\caption{Scheduler for Subcircuit Placement}
\begin{algorithmic}[1]
\STATE Input: 
\item[] $S_{EPR} = {s_0,s_1,...,(s_i,s_j),...}$ : subcircuits including pairs for EPR connection
\item[] $W=\{w_0:N_0, w_1:N_1,...,w_n:N_n\}$: Noise model $N_i$ in worker $w_i$
\FOR{$S_{EPR}$}
    \STATE $p, q \gets CG(S)$
    \FOR{$W$}
        \STATE $ w_i \gets (N_i, p, q)$
    \ENDFOR
    \STATE $EC \gets [S_j: w_i]$
\ENDFOR
\RETURN $EC$
\end{algorithmic}
\label{alg:3}
\end{algorithm}

\subsubsection{Subcircuit Scheduling}
Once we have the subcircuits, and some of them contain the EPR, the Algorithm~\ref{alg:3} schedule them into the system for best quantum worker pairs that is used to execute them. It accepts subcircuits $S_{EPR}$, and the Noise model in each worker $W=\{w_0:N_0,...,w_n:N_n\}$. The algorithm begins by iterating through the set of subcircuits $S_{EPR}$(Line 2-8). For each subcircuit, it computes single qubit gate $p_i$ and two qubits gate $q_i$ using the function \textbf{CG(S)} calculate the number of two types of gate based on the worker's noise model(Line 3). Then, for each worker $W_i$, analysis the affection of noise $N_i$ to $S_i$ with $p$ and $q$. It finds the specific worker $W_i$ for subcircuit $S_j$. This process repeats for all workers(Line 4-6). After finding the best optimal assignment, it update $[S_j: w_i]$ to \textbf{EC}, that is the subcirucit placement with subcircuit $s_j$ in worker $w_i$  (Line 7), and return the final scheduling plan that with the minimize total sampling overhead for whole multi-node system from equation~\ref{eq:epr_overhead}.


\begin{algorithm}
\caption{\sol~ system optimization}
\begin{algorithmic}[1]
\STATE Input:  $C$, $W$, $N$, $E$
\STATE $S_{EPR} \gets Algorithm~\ref{alg:2}($C$, $W$, $N$, $E$)$
\STATE $EC \gets Algorithm~\ref{alg:3}(S_{EPR},W)$
\FORALL{$EC$}
    \STATE $RC \gets$ Execute $S_j$ on $w_i$
\ENDFOR
\STATE $R \gets$ Reconstruct $RC$
\RETURN $R$
\end{algorithmic}
\label{alg:4}
\end{algorithm}

Algorithm ~\ref{alg:4} use both algorithms ~\ref{alg:1}-\ref{alg:3} then execute all subcircuits to get all subcircuit results, denoted as \textbf{RC}, and use them to reconstruct the final result, denoted as \textbf{R} of the original circuit, a fully distributed probability.
\section{Evaluation}


In this section, we evaluate the performance of \sol, focusing on two key metrics: fidelity and system expenditure for circuit cutting and EPR pairs. We conducted experiments using real quantum hardware, quantum emulators, and simulators with noise models. First, we present the details of our implementation, followed by the workload description, and finally the results from our experiments.

\subsection{Implementation }


We implemented the \sol~system using Python 3.10, integrating Qiskit 1.2, a leading quantum programming library provided by IBM Quantum, for superconducting quantum platforms, and Pytket 1.32.0, a quantum SDK provided by Quantinuum, for trapped-ion quantum platforms.


Our experiments were conducted on three IBM Quantum machines, \textit{IBM\_Brisbane}, \textit{IBM\_Kyiv}, and \textit{IBM\_Sherbrooke}, each with 127 qubits. Additionally, we utilized IBM-Q noisy emulators, specifically AerSimulator with fake backends simulating noisy environments built on real quantum hardware (e.g., Auckland, TorontoV2, and MontrealV2). Beyond IBM-Q, we also evaluated \sol~ on the Quantinuum Nexus platform with the H1-Emulator. For classical processing, we leveraged the Google Cloud Platform using its e2-highmem-16 instance equipped with AMD Rome x86/64 processors.



\textbf{Experiment Settings:} \sol~ considers a multi-node quantum system consisting of multiple workers, each with a specific qubit count. To assess performance under different configurations, we varied the number of workers from 3 to 7, with qubit counts of 15, 20, and 25 per worker. Depending on the number of workers, there were between 1 and 6 EPR pairs connecting different nodes. To obtain the fidelity values on 127-qubit real quantum machines, we set a large number for shots due to the machine, 40,960 for {IBM\_Brisbane}, and circuit sizes.

\subsection{Workloads, Baselines, and Metrics}

{\bf Workloads}: The circuits evaluated in our experiments include a range of well-known quantum algorithms and randomly generated circuits with varying structures and complexities. Specifically, we included:

\begin{itemize}
    \item  Bernstein-Vazirani (BV) Algorithm: A quantum algorithm used to determine a hidden bit string~\cite{bernstein1993quantum}.
\item ADDER: A quantum circuit for performing binary addition~\cite{zhang2023characterizing}.
\item Hardware Efficient Ansatz (HWEA): A circuit typically used in variational algorithms~\cite{kandala2017hardware}.
\item Supremacy Circuit: A random circuit designed to demonstrate quantum supremacy~\cite{boixo2018characterizing}. Google utilizes this type of circuit to demonstrate a quantum advantage in ~\cite{arute2019quantum}.
\end{itemize}

These circuits were created with different qubit counts, ranging from 20 to 100, to evaluate the system's scalability under various conditions.


{\noindent\bf Baselines}: We compare the performance of \sol~ with IBM Qiskit-Addon-Cut, the state-of-the-art solution for circuit cutting, which serves as the baseline for our comparison. However, Qiskit-Addon-Cut does not support multi-node environments (i.e., generating EPR pairs) and lacks noise-aware algorithms for circuit analysis and subcircuit scheduling. To address this limitation, we extend the Qiskit-Addon-Cut by introducing random EPR pairs. Specifically, we randomly select subcircuit pairs \((s_i, s_j)\) and assign them to random worker pairs \((w_i, w_j)\). We refer to this solution as Qiskit-Random-\textit{x}, where \textit{x} represents the number of EPR pairs.

{\noindent\bf Metrics}: In our experiments, we considered the following three metrics: (1) Fidelity; It measures how closely the actual quantum state matches the ideal or target quantum state, which is critical for evaluating quantum applications. (2)  System Expenditure in a Multi-node Environment (SEME): As described in Equation~\ref{eq:mse}, SEME measures the overhead introduced by both circuit cutting and the use of EPR pairs. In Figures~\ref{fig:cost of epr} and~\ref{fig:workers}, SEME demonstrates the percentage comparison with Qiskit-Addon-Cut.

\subsection{Fidelity}


In this subsection, we present our fidelity results under two different settings and environments:


\begin{itemize}
    \item \textbf{Single-Node}: This setting involves real quantum machines or individual emulators (i.e., IBM Qiskit fake providers). Due to limitations in quantum hardware and restricted access to on-board EPR pairs, the Single-Node environment utilizes either one real quantum machine or one emulator for the experiments. All subcircuits are executed on a single node. For Qiskit-Addon-Cut, which does not utilize EPR pairs, the fidelity values are computed without considering EPR success rates. For all other solutions, fidelity is computed with an EPR success rate ($SR$) of 0.9, which is the lowest value observed for short-distance or localized quantum systems (e.g., superconducting qubits, trapped ions, and neutral atoms). In this scenario, Algorithms 1-4 are not applicable.
    
    \item \textbf{Multi-Node}: In this setting, we simulate a multi-node system using multiple emulators in the cloud. Various IBM Qiskit fake providers are deployed as backends, each with a different noise model based on data from real quantum machines. In this environment, Algorithms 1-4 are fully operational and utilized.
\end{itemize}

\begin{figure*}[htbp]
    \centering
    \includegraphics[width=1\linewidth]{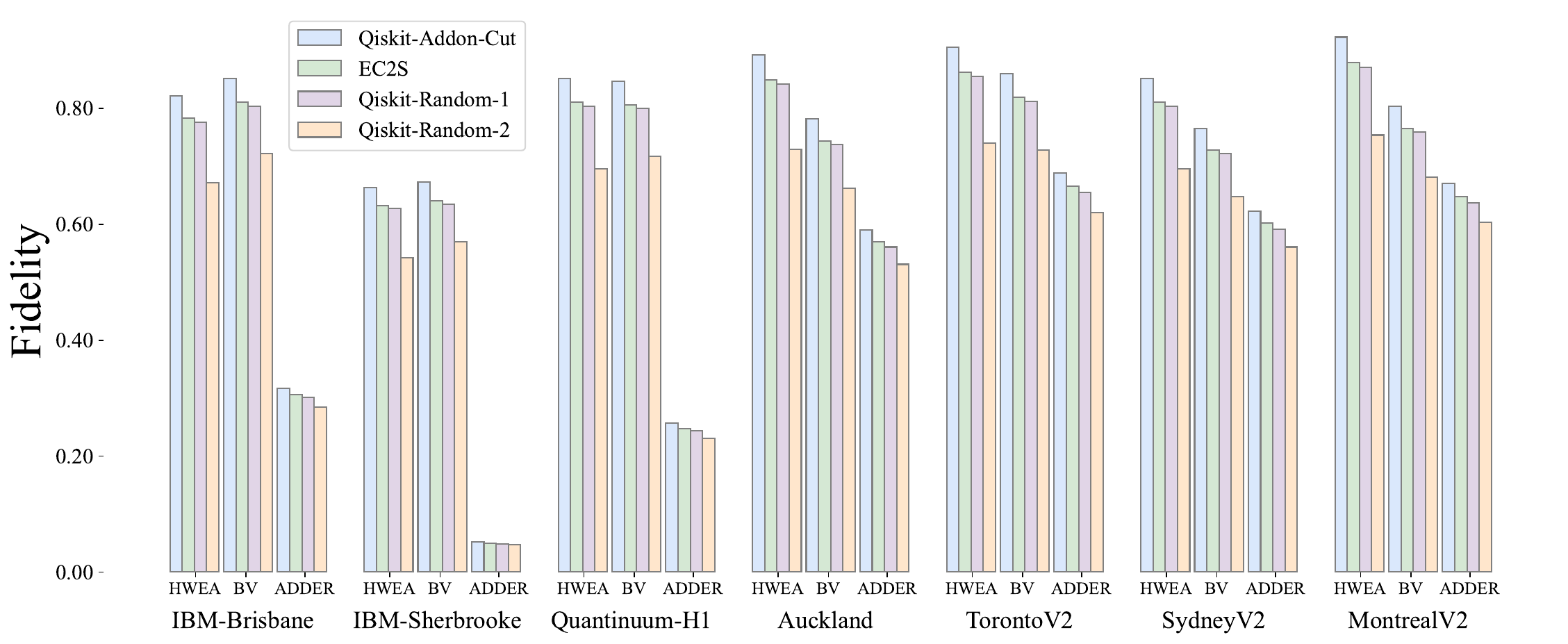}
    \caption{Fidelity Results in a Single-Node Environment}
    \label{fig:real_F}
\end{figure*}

{\noindent \bf Single-Node Results}: Figure~\ref{fig:real_F} shows the results for the single-node setting with $SR = 0.9$. Among the experiments, \textit{IBM\_Brisbane} and \textit{IBM\_Sherbrooke} are real quantum machines, while the others are emulators. Notably, H1 is an emulator from Quantinuum, which simulates a trapped-ion platform. We used a 25-qubit HWEA, a 25-qubit BV algorithm, and a 24-qubit ADDER circuit as our workloads, with the qubit count capped at 20.

Qiskit-Addon-Cut consistently achieves the highest fidelity across all solutions. This is because, unlike other approaches, it does not use EPR pairs, and all subcircuits run on the same machine in the Single-Node environment. Therefore, it is not affected by the EPR success rate ($SR = 0.9$). On the other hand, \sol~ and Qiskit-Random-1/2 are influenced by this success rate. Due to the EPR success rate, subcircuits require more shots for execution, and noise accumulates over longer runs. This accumulation results in lower fidelity for these solutions. This effect is evident as Qiskit-Random-2 exhibits lower fidelity than Qiskit-Random-1 due to the additional EPR pair in use. Furthermore, we observe that the ADDER circuit consistently performs the worst among the four circuits tested. This can be attributed to the higher gate count and circuit depth in the ADDER circuit. As circuit depth increases, subcircuits become more noise-resistant, leading to lower fidelity.

While Algorithms 1-4 are not applicable in this environment, Qiskit-Addon-Cut provides an optimal fidelity that we can possibly achieve.

\begin{figure}[htbp]
\centering
\includegraphics[width=1\linewidth]{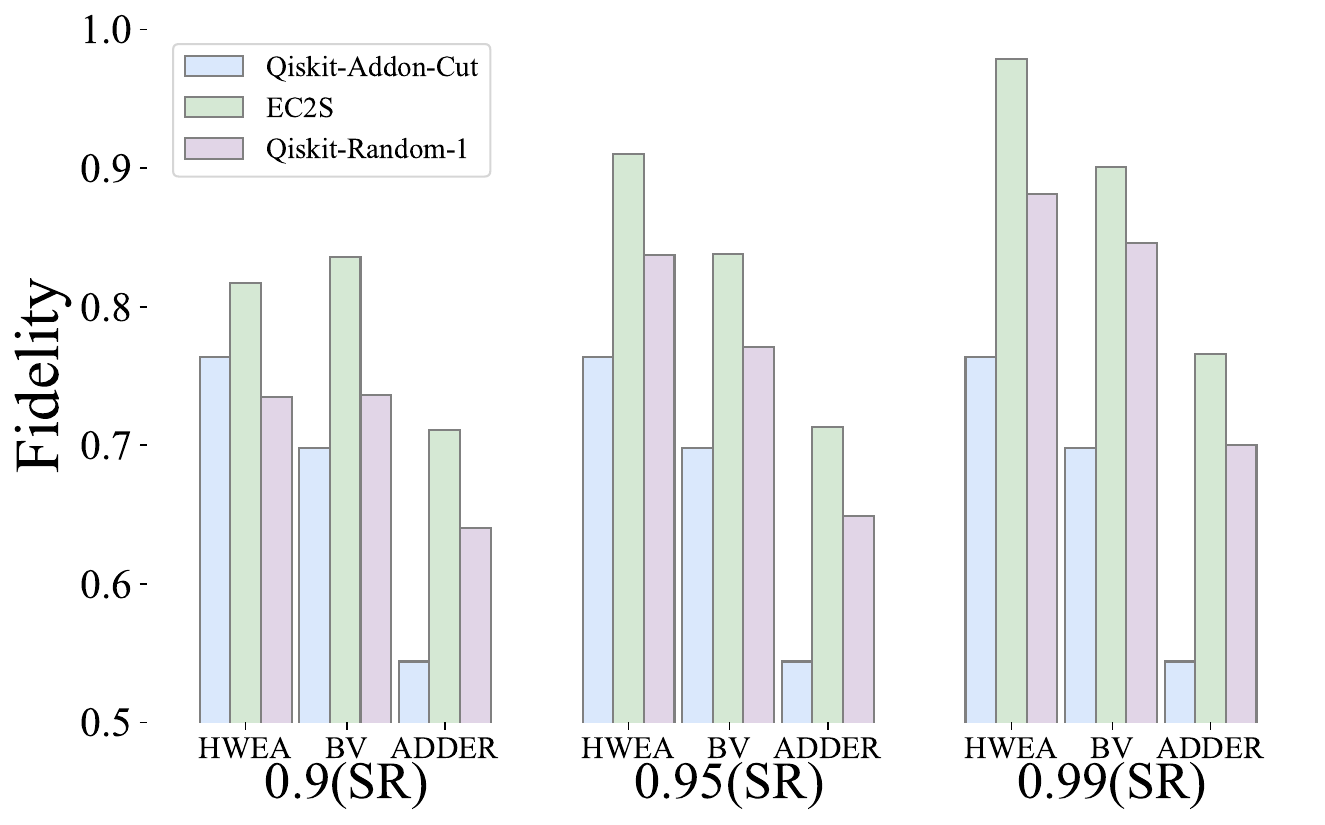}
\caption{Fidelity Results in a Multi-Node Environment: 4 workers randomly selected from the pool    of Auckland, Toronto, Sydney, and Montreal (with repetitions)}
\label{fig:multi-node}
\end{figure}

\noindent {\bf Multi-Node Results}: Figure~\ref{fig:multi-node} presents the fidelity results from our multi-node experiments. \sol~ builds a 4-worker system by randomly selecting emulator backends from a pool consisting of Auckland, Toronto, Sydney, and Montreal. In this scenario, \sol~ is fully functional, with all algorithms in effect. In addition to $SR = 0.9$, we evaluate the system under two other success rates, $SR = 0.95$ and $SR = 0.99$. Unlike the trend observed in Figure~\ref{fig:real_F}, Qiskit-Addon-Cut is no longer the best performer. \sol~ consistently outperforms it. For example, at $SR = 0.9$, ~\sol~ achieves fidelity improvements of 5.3\%, 12.8\%, and 16.7\% for HWEA, BV, and ADDER, respectively. The significant improvement in ADDER is due to its higher number of gates compared to HWEA and BV, making it more susceptible to worker noise. Our ~\sol~optimization effectively reduces the impact of the noise model on subcircuits.
As $SR$ increases from 0.9 to 0.99, \sol~fidelity increases 16.2\%, 6.5\%, and 5.5\% for HWEA, BV, and ADDER. At $SR = 0.99$, the circuit Adder achieves the best improvement of fidelity from 54.4\% to 76.6\%, which increases by 40.8\%. This improvement is due to \sol's ability to leverage EPR pairs, allowing it to take advantage of multiple workers with different noise levels. \sol~ selects the optimal workers for executing subcircuits with the consideration of noises. In contrast, Qiskit-Addon-Cut is restricted to a single worker due to its inability to utilize EPR pairs. 

Compared to Figure~\ref{fig:multi-node} at $SR = 0.9$ and Figure~\ref{fig:real_F} where the system transitions from a single-node to a multi-node setup, \sol~ optimizes performance by selecting physical EPR pairs across workers with varying noise levels for logical EPR connections. For example, HWEA's fidelity decreases in the range of $[3.2\%, 4.7\%]$ in the single-node configuration, while it increases by 5.3\% in the multi-node configuration. This demonstrates the effectiveness of \sol's optimization. Also, this applies to the other two types of circuit BV and ADDER, their fidelity become from $-[3.2\%, 4.1\%]$ and $-[0.2\%, 2.3\%]$ to $+13.8\%$ and $+16.7\%$, respectively.


Qiskit-Random-1 also benefits from multiple workers by using one EPR pair, but it records lower fidelity compared to \sol. For example, the improvement of \sol~ range is $[7,1\%, 10\%]$ for four HWEA, BV, and ADDER. Circuit BV at $SR = 0.9$, fidelity of Qiskit-Random-1 is 73.6\%, whereas \sol~achieves 83.6\%, representing a 10\% improvement. This difference is due to \sol~ not only utilizing EPR pairs but also incorporating noise-aware algorithms to select the best subcircuit pairs and assign them to the most suitable worker pairs, that is applying the logical EPR pair to physical EPR pair, considering noise levels. For instance, BV is cut into 4 subcircuit ${s_0,s_1,s_2,s_3}$, \sol~find logical subcircuit pair is $(s_0,s_1)$ to place on worker $(w1,w2)$ that are connected by physical EPR pair, and the $s_2$ in $w_3$, $s_3$ in $w_0$. For Qiskit-Random-1, the placement is $s_0$ in $w_0$,$s_1$ in $w_1$,$s_2$ in $w_2$,$s_3$ in $w_3$. The EPR is generated between $s_0, s_1$ on $w_0, w_1$.

\subsection{SEME and Number of EPR pairs}

\begin{figure}[htbp]
    \centering
    \includegraphics[width=1\linewidth]{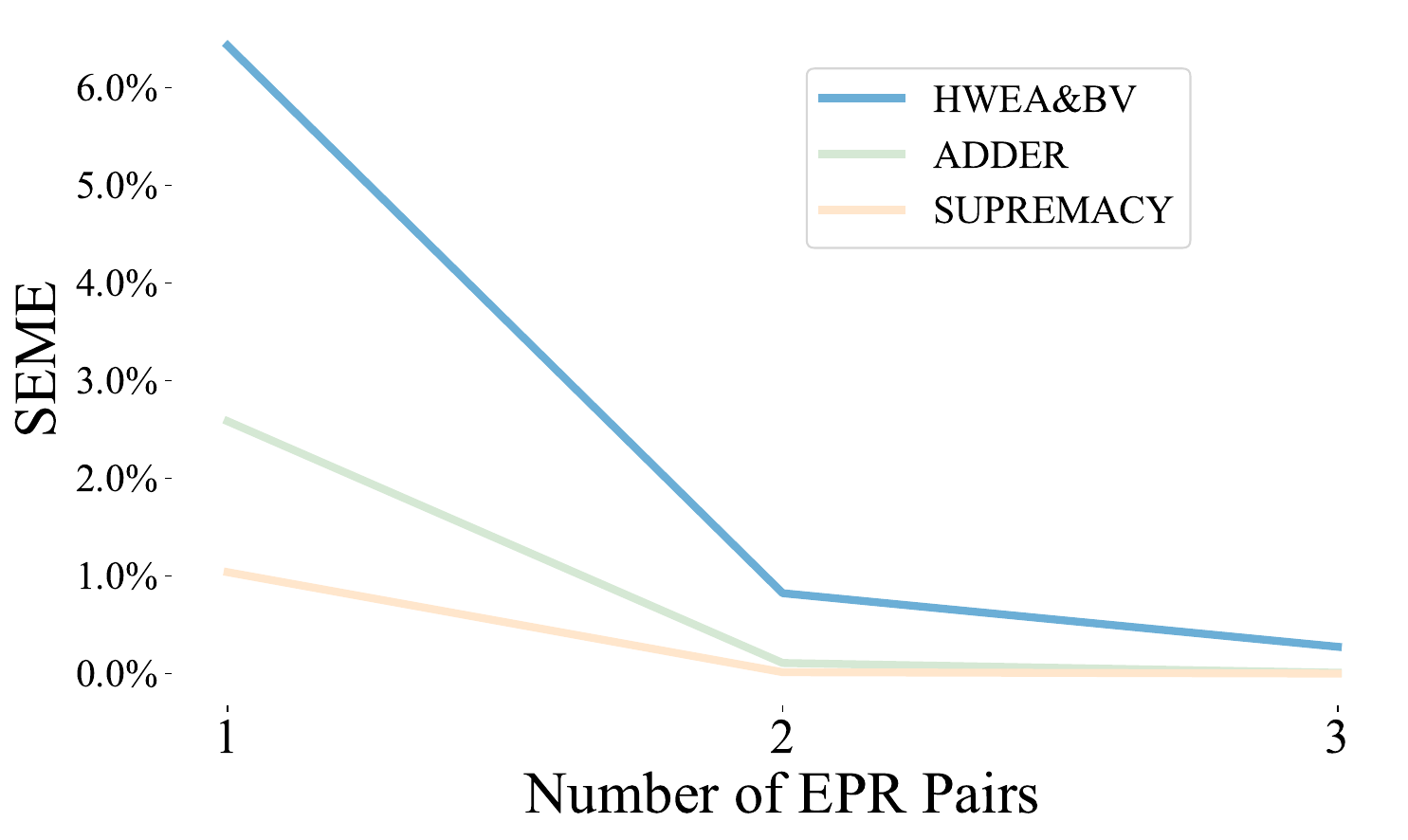}
    \caption{SEME Cost in a 4-worker Multi-Node Environment with 1, 2, and 3 EPR Pairs}
    \label{fig:cost of epr}
\end{figure}

\begin{figure}[htbp]
    \centering
    \includegraphics[width=1\linewidth]{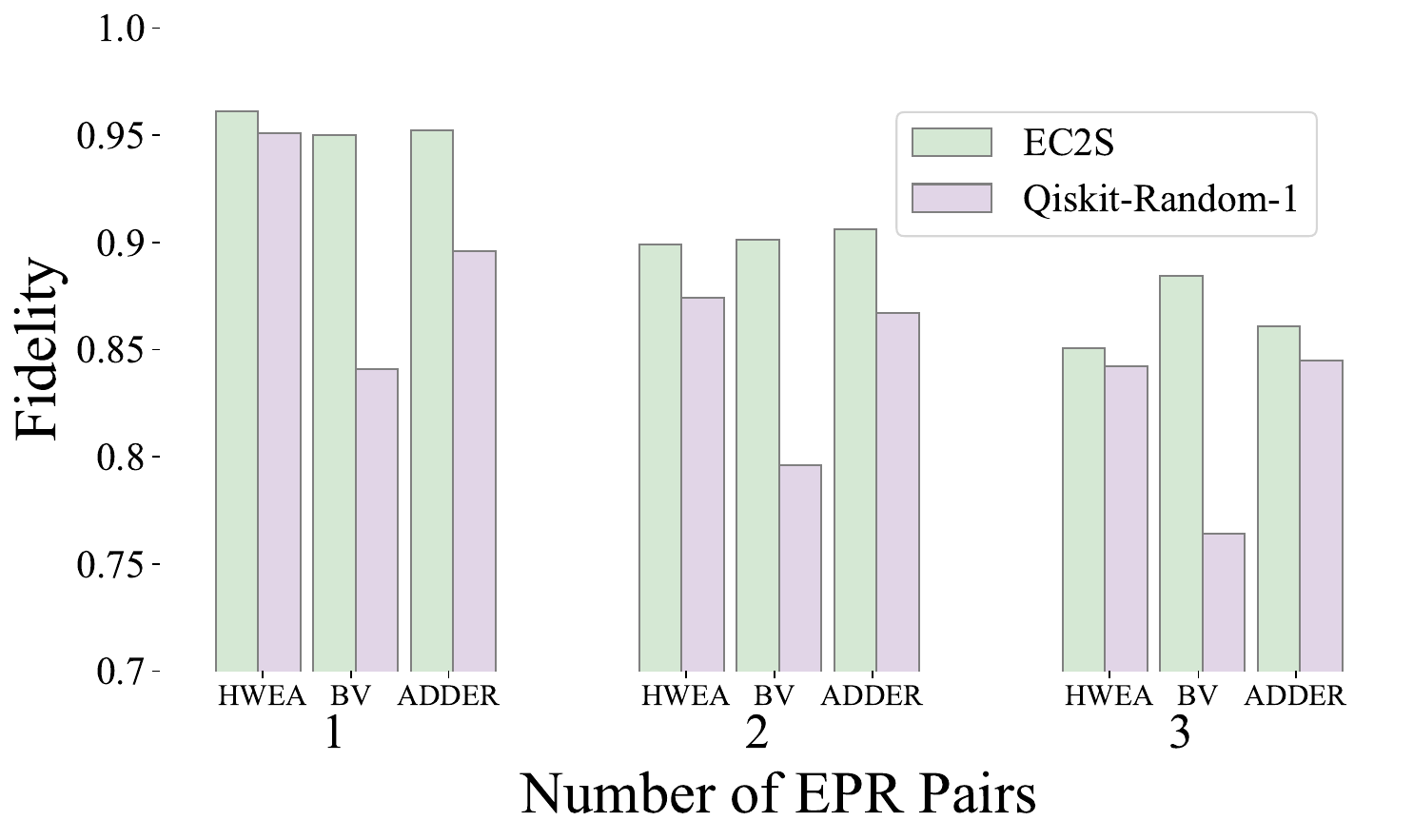}
    \caption{Fidelity in a 4-worker Multi-Node Environment with 1, 2, and 3 EPR Pairs}
    \label{fig:fidelity of epr}
\end{figure}

Figure ~\ref{fig:cost of epr} and Figure~\ref{fig:fidelity of epr} are results for evaluating the same circuit. They illustrate the percentage trend of SEME costs in comparison to the Qiskit-Addon-Cut, as shown in the Figure ~\ref{fig:cost of epr} and fidelity in Figure~\ref{fig:fidelity of epr} as the number of EPR pairs increases for the same circuit. The results show that as the number of EPR pairs increases, the cost decreases, but fidelity also decreases.


We use the SEME value from Equation~\ref{eq:SEME} of Qiskit-addon-cut as a reference (set to \textbf{1} unit), representing an EPR-free result. As the number of EPR pairs increases to 1, the SEME cost for HWEA and BV decreases from 100.0\% to 6.4\% by applying one EPR pair and optimizing the scheduler, and continues to decrease to 0.8\%. However, fidelity decreases from 96.1\% to 89.9\% as the number of EPR pairs increases from 1 to 2. With a further increase to three EPR pairs, the SEME cost becomes 0.2\%, while fidelity decreases to 85.1\%. As a result, we observe that increasing the number of EPR pairs from 1 to 3 reduces the HWEA's SEME cost from 6.4\% to 1.0\%, meanwhile, the fidelity decreases from 96.1\% to 85.1\%. Although increasing the number of EPR pairs can further reduce SEME, we aim to maintain higher fidelity, highlighting the trade-off involved in selecting more EPR pairs.



\subsection{SEME and Number of Workers}
In the \sol~system, the number of workers plays a critical role in determining the efficiency and cost of performing various tasks, including subcircuit execution and reconstruction processes. As the number of workers in a distributed system increases, the number of subcircuits that can be cut increases, providing more flexibility for finding logical EPR pairs during the cutting process and selecting physical EPR pairs in a multi-node system. Although a more complex system configuration can increase the time required to find cuts, it ultimately reduces the overall cost.

\begin{figure}[htbp]
    \centering
    \includegraphics[width=1\linewidth]{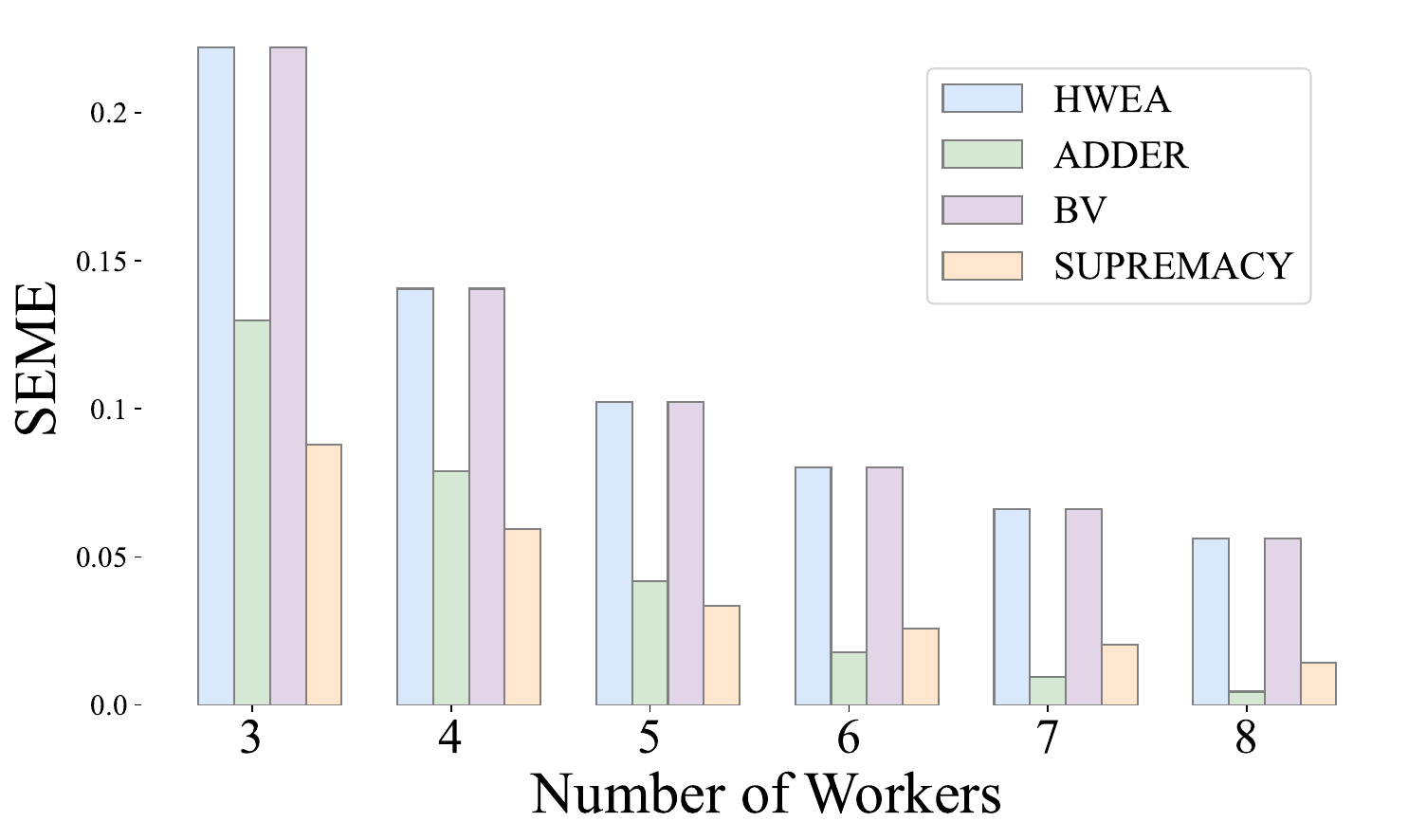}
    \caption{SEME on different number of Workers}
    \label{fig:workers}
\end{figure}

In Figure ~\ref{fig:workers}, we show the percentage of SEME cost for the HWEA circuit (100-qubits). With Qiskit-addon-cut, the result is 100.0\%, while \sol~ reduces SEME cost from 22.2\% with 3 workers to 5.6\% with 8 workers. For ADDER (70-qubits) and Supremacy (64-qubits) circuit, the SEME decreases from 13.0\% to 0.5\% and 8.8\% to 1.4\%, respectively, reducing the expenditure by up to 99.5\%.

\section{Conclusion}

In this paper, we aim to reduce the significant overhead associated with traditional circuit-cutting techniques. We consider a multi-node quantum-classical system in which quantum workers can be connected via dynamically generated EPR pairs. By leveraging EPR pairs, we connect subcircuits produced through circuit cutting, while accounting for the impact of noise. Based on this, we propose the \sol~ system, which enhances the efficiency of circuit cutting by optimizing both the cutting and scheduling processes.
Specifically, \sol~ selects the optimal subcircuit pairs that aim to reduce overhead and connects them with logical EPR pairs. Furthermore, \sol~ optimizes the selection of quantum worker pairs for the allocation of physical EPR pairs, taking into account the overhead introduced by the EPR connections. Finally, \sol~ schedules subcircuits to worker pairs by evaluating the noise profiles of different workers. This noise-aware approach minimizes the number of subcircuits in the reconstruction phase, thereby reducing computational overhead. 
The intensive experiments demonstrate significant improvement, with up to 16.7\% fidelity and 40.8\% increment. The SEME computational overhead cost also significant reduce to as little as 0.5\%.

\bibliographystyle{ACM-Reference-Format}

\bibliography{references}

\end{document}